\documentclass[12pt,preprint]{aastex}
\usepackage{natbib}
\usepackage{amssymb,amsmath}
\usepackage[colorlinks,
            linkcolor=blue,
            anchorcolor=blue,
            citecolor=blue,
            ]{hyperref}
\usepackage{lineno}
\begin{document}


\title{Gamma-Ray Emission Produced by $r$-process Elements from Neutron Star Mergers}
\author{Meng-Hua Chen$^{1}$, Li-Xin Li$^{2*}$, Da-Bin Lin$^{1}$, En-Wei Liang$^{1*}$}
\altaffiltext{1}{Guangxi Key Laboratory for Relativistic Astrophysics, School of Physical Science and Technology, Guangxi University, Nanning 530004, China; lew@gxu.edu.cn}
\altaffiltext{2}{Kavli Institute for Astronomy and Astrophysics, Peking University, Beijing 100871, China; lxl@pku.edu.cn}
\begin{abstract}
The observation of a radioactively powered kilonova AT~2017gfo associated with the gravitational wave-event GW170817 from binary neutron star merger proves that these events are ideal sites for the production of heavy $r$-process elements.
The gamma-ray photons produced by the radioactive decay of heavy elements are unique probes for the detailed nuclide compositions.
Basing on the detailed $r$-process nucleosynthesis calculations and considering radiative transport calculations for the gamma-rays in different shells, we study the gamma-ray emission in a merger ejecta on a timescale of a few days.
It is found that the total gamma-ray energy generation rate evolution is roughly depicted as $\dot{E}\propto t^{-1.3}$.
For the dynamical ejecta with a low electron fraction ($Y_{\rm e}\lesssim0.20$), the dominant contributors of gamma-ray energy are the nuclides around the second $r$-process peak ($A\sim130$), and the decay chain of $^{132}$Te ($t_{1/2}=3.21$~days) $\rightarrow$ $^{132}$I ($t_{1/2}=0.10$~days) $\rightarrow$ $^{132}$Xe produces gamma-ray lines at $228$~keV, $668$~keV, and $773$~keV.
For the case of a wind ejecta with $Y_{\rm e}\gtrsim0.30$, the dominant contributors of gamma-ray energy are the nuclides around the first $r$-process peak ($A\sim80$), and the decay chain of $^{72}$Zn ($t_{1/2}=1.93$~days) $\rightarrow$ $^{72}$Ga ($t_{1/2}=0.59$~days) $\rightarrow$ $^{72}$Ge produces gamma-ray lines at $145$~keV, $834$~keV, $2202$~keV, and $2508$~keV.
The peak fluxes of these lines are $10^{-9}\sim 10^{-7}$~ph~cm$^{-2}$ s$^{-1}$, which are marginally detectable with the next-generation MeV gamma-ray detector \emph{ETCC} if the source is at a distance of $40$~Mpc.
\end{abstract}

\keywords{gamma-ray burst: general -- gravitational waves -- nuclear reactions, nucleosynthesis, abundances -- stars: neutron}

\section{Introduction}
\label{sec1}

The rapid neutron capture process ($r$-process) is believed to be responsible for the production of about half of the elements heavier than iron in our
universe (\citealp{Burbidge1957,Cameron1957}; see \citealp{Cowan2021} for a recent review).
Binary neutron star (or neutron star and black hole) mergers have long been considered as promising sites of the $r$-process nucleosynthesis \citep{Lattimer1974,Lattimer1976,Symbalisty1982}.
Hydrodynamic simulations of binary neutron star mergers reveal that a small amount of mass ($\sim10^{-4}-10^{-2}M_\odot$) with a low electron fraction ($Y_{\rm e}\sim0.1-0.4$) is ejected with subrelativistic velocities ($\sim0.1-0.3c$) (see \citealp{Shibata2019} for a recent review).
Merger ejecta would be an ideal site for the $r$-process nucleosynthesis.
Radioactive decay of $r$-process elements freshly synthesized in the merger ejecta is expected to produce an ultraviolet/optical/near-infrared transient \citep{Li1998,Metzger2010,Korobkin2012,Barnes2013,Kasen2013,Tanaka2013,Metzger2014,Barnes2016},
which is called a ``kilonova'' (see \citealp{Metzger2019} for a recent review).
Our understanding of the $r$-process advanced dramatically after the discovery of the first neutron star merger event GW170817 \citep{Abbott2017a}.
Approximately eleven hours after the merger, the electromagnetic transient, named AT~2017gfo, was observed in the ultraviolet, optical and near infrared wavelengths in the galaxy NGC~4993 \citep{Abbott2017b,Arcavi2017,Chornock2017,Coulter2017,Cowperthwaite2017,Drout2017,Evans2017,Kasliwal2017,Lipunov2017,McCully2017,Nicholl2017,Pian2017,Smartt2017,
Soares-Santos2017,Tanvir2017,Valenti2017}.
The observed features are broadly consistent with a kilonova model \citep{Kasen2017,Tanaka2017}, indicating that $r$-process elements have been synthesized in this event.
The merger event provides the first strong evidence for an astrophysical site of $r$-process elements production.

Observations of the kilonova transient may be used to estimate the masses and detailed nuclide compositions of the merger ejecta.
However, the ejecta mass estimate involves many systematic uncertainties, mostly due to the uncertain opacity, and to lesser extent due to the heating rate, thermalization efficiency and ejecta properties \citep{Kasen2013,Tanaka2013,Barnes2016,Kawaguchi2018,Wollaeger2018}.
The opacity of the merger ejecta is very sensitive to the abundance pattern of heavy elements.
Thus, even with the pristine data of GW170817, it is difficult to accurately estimate the ejecta mass.
In addition, all of these studies assumed that the radioactive decay powers the kilonova transient, but the later red emissions may also arise from delayed energy injection from a long-lived remnant neutron star \citep{Yu2018,Ren2019}.
With all these uncertainties,
it is difficult to accurately estimate the ejecta mass based solely on the kilonova lightcurve. Obtaining detailed nuclide compositions of the merger ejecta is even more challenging.

The gamma-ray photons produced by the radioactive decay of $r$-process elements provide a unique opportunity to probe the detailed compositions of the merger ejecta.
According to theoretical estimates, the ejecta will become optically thin after a day to a few days caused by the subrelativistic expansion \citep{Pian2017,Drout2017,Troja2017,Kilpatrick2017,Shappee2017,Waxman2018}.
Then, the gamma-ray photons would escape from the ejecta material directly.
Detection and observation of these radioactive gamma-ray emission would be the best approach for directly probing the detailed yields of heavy elements.
In nearby supernovae, the gamma-ray emission produced by radioactive nuclides has been detected, e.g., the gamma-ray lines of $^{56}$Co from the Type~II~SN~1987A \citep{Matz1988,Teegarden1989}, and those of $^{56}$Ni and $^{56}$Co from the Type~Ia~SN~2014J \citep{Churazov2014,Diehl2014}.
The two gamma-ray emission lines of $^{56}$Co at energies $847$ and $1238$~keV from the SN~2014J in the nearby galaxy M82 were detected by \emph{INTEGRAL}.
From the observed luminosity of the gamma-ray lines, it is successfully derived that about 0.6$M_{\odot}$ radioactive $^{56}$Ni were synthesized during the explosion \citep{Churazov2014}.
The detection of radioactive gamma-ray emission can provide a conclusive evidence for identification specific heavy elements, in principle.
Hence, it is worthwhile to study the spectrum and luminosity of the gamma-ray emission in a neutron star merger.

Gamma-ray emission from neutron star merger through the decays of $r$-process elements is a timely topic which is important for both kilonova and $r$-process study.
There have been several groups working on this topic.
\citet{Hotokezaka2016} studied the gamma-ray emission from neutron star merger with a dynamic $r$-process network and found that it could be directly observed from an nearby event ($\leq3-10$~Mpc) with future gamma-ray detectors.
\citet{Li2019} studied the features of the gamma-ray emission from a neutron star merger in detail.
Instead of using $r$-process network, he used a Monte Carlo method to generate the initial abundance of unstable nuclides and compared the gamma-ray spectra produced by neutron-rich nuclides and proton-rich nuclides.
\citet{Wu2019b} studied the gamma-ray line signals from long-lived nuclei to search for remnants of past neutron star mergers in our Galaxy.
\citet{Korobkin2020} studied the gamma-ray emission both in the kilonova phase and in the remnant epoch, with a 3D radiative transport code and the $r$-process network.
\citet{Wang2020} included the contribution from fission fragments to calculate the high energy gamma-ray emission from neutron star mergers.
\citet{Lapuente2020} studied the contribution of the kilonova gamma-rays to the diffusive gamma-ray background emission.
In the present work, based on the detailed $r$-process nucleosynthesis calculations and a multi-shell model for radiative transport, we will calculate the shape and feature of the gamma-ray spectra in the merger ejecta,
identifying the features in the emission spectrum associated with $r$-process elements.

The paper is organized as follows. In Section~\ref{sec2}, we describe the procedure to calculate gamma-ray emission.
In Section~\ref{sec3}, we present the results of radioactive gamma-ray emission.
Section~\ref{sec4} contains the summary and discussion.

\section{Procedure to Calculate Gamma-Ray Emission}
\label{sec2}

To calculate the gamma-ray emission from $r$-process nucleosynthesis,
one should know the species of nuclide and the corresponding abundances inside the merger ejecta.
In this work, the $r$-process nucleosynthesis in the merger ejecta is performed
based on the code of \emph{SkyNet} \citep{Lippuner2017},
which evolves the abundances of nuclides under the influence of nuclear reactions,
involving $7843$ nuclides ranging from free neutrons and protons to $^{337}$Cn($Z=112$) and
including more than $1.4\times10^5$ nuclear reactions.
The nuclear masses and partition functions used in \emph{SkyNet} are
taken from REACLIB \citep{Cyburt2010}.
The initial conditions of ejecta are set as followed.
The dynamical ejecta is calculated with entropy $s=10~k_B$/baryon, expansion timescale $\tau_{\rm dyn}=10$~ms, and electron fraction $Y_{\rm e}\sim0.05-0.20$.
For the wind ejecta, we choose entropy $s=20~k_B$/baryon, expansion timescale $\tau_{\rm dyn}=30$~ms, and electron fraction $Y_{\rm e}\sim0.25-0.40$.
The adopted initial conditions of the ejecta are consistent with those found in the numerical simulations carried out by \cite{Nedora2021}.

The total gamma-ray energy generation is given by summing the energy generation of all decay mode for all nuclides. We denote a nuclide species by an index $i$, and a decay mode by an index $j$.
Then, the gamma-ray energy generation is related to the time evolution of the nuclide abundance.
Following \cite{Lippuner2017}, the abundance of nuclides can be defined as
\begin{equation}
Y_i=\frac{N_i}{N_B},
\end{equation}
where $N_i$ and $N_B$ are the total numbers of particles of $i$th nuclide species and baryons, respectively.
Then the gamma-ray energy generation rate can be written as
\begin{equation}
\dot{E_{\gamma}}=N_B\sum_iY_i\sum_j\frac{\varepsilon_{ij}}{\tau_{ij}}.
\end{equation}
Here, $\tau_{ij}$ is the mean lifetime of a nuclide and estimated with $\tau_{ij}=t_{1/2,ij}/\ln{2}$,
where $t_{1/2,ij}$ is the half-life of the nuclide and obtained by $t_{1/2,ij}=t_{1/2,i}/B_{ij}$
with $B_{ij}$ being the branching ratio of a nuclide in a given decay mode.
The $\varepsilon_{ij}$ is the total energy of the gamma-rays generated in the $j$th decay mode of the $i$th nuclide and can be written as
\begin{equation}
\varepsilon_{ij}=B_{ij}\sum_kh_{ijk}\epsilon_{ijk},
\end{equation}
where $\epsilon_{ijk}$ is the $k$th photon energy of the gamma-ray generated in the $j$th decay mode of the $i$th nuclide and $h_{ijk}$ is the corresponding intensity (probability of emitting gamma-ray photon per decay).
In our calculations, the gamma-ray radiation data for the unstable nuclides in each decay mode (including $\alpha$ decay and $\beta$ decay) are taken from the NuDat2 database\footnote{http://www.nndc.bnl.gov/nudat2/}.

To calculate the spectrum of the gamma-ray emission, we divide the photon energy range of $[10^0, 10^4]$~keV into $400$ energy bins in the logarithmic space.
The specific photon energy rate in a bin of photon energy, e.g., [$\epsilon_1,\epsilon_2$], is defined by
\begin{equation}
L(\epsilon)~d\epsilon=\sum_{\epsilon_1}^{\epsilon_2} N_B \sum_i Y_i\sum_j \frac{B_{ij}}{\bar{\tau}_{ij}}\sum_k h_{ijk}\epsilon_{ijk}.
\end{equation}
Then, one can have the emission coefficient $j_{\epsilon}=L(\epsilon)/(4\pi V)$, which is used in our calculation of the observed photon flux, where $V$ is the volume of the merger ejecta.

To obtain the emitted gamma-rays from the ejecta, we use the radiative transfer equation to get the intensity $I_{\epsilon}$, i.e.,
\begin{equation}
\label{eq:tran}
\frac{dI_{\epsilon}}{dl}=-\alpha_{\epsilon} I_{\epsilon}+j_{\epsilon},
\end{equation}
where the absorption coefficient $\alpha_{\epsilon}$ depends on the mass density $\rho_{\rm ej}$ and the opacity $\kappa(\epsilon)$ as $\alpha_{\epsilon}=\rho_{\rm ej}\kappa(\epsilon)$,
and $l$ is the photon path length. The optical depth $\tau_{\epsilon}(l)$ for the photons is $\tau_{\epsilon}(l)=\int \rho_{\rm ej}\kappa(\epsilon)dl$. Thus, the formal solution of Equation~\ref{eq:tran} for photons traveling from $l_0$ to $l_m$ is written as 
\begin{equation}
I_{\epsilon}=I_{\epsilon}(l_0)e^{(-\tau_{\epsilon}(l_m)-\tau_{\epsilon}(l_0))}
+\int_{l_0}^{l_m}j_{\epsilon}e^{(-\tau_{\epsilon}(l_m)-\tau_{\epsilon}(l))}dl.
\end{equation}

Similar to the work of \cite{Metzger2019}, a spherical expanding ejecta is assumed in this work.
The merger ejecta is divided into $N$ shells with different expansion velocity $v_n(1\leq n\leq N)$, where $v_1=v_{\rm min}$, $v_N=v_{\rm max}$, and $N=100$.
The mass of each shell is determined by the density distribution, which is taken as a power law (\citealp{Nagakura2014})
\begin{equation}
\rho_{\rm ej}(R_n)=\frac{(\delta-3)M_{\rm ej}}{4\pi R_{\rm max}^3}\left[\left(\frac{R_{\rm min}}{R_{\rm max}}\right)^{3-\delta}-1\right]^{-1}\left(\frac{R_n}{R_{\rm max}}\right)^{-\delta},
\end{equation}
where $M_{\rm ej}$ is the total mass of the ejecta, and $R_{\rm max}=v_{\rm max}t$ and $R_{\rm min}=v_{\rm min}t$ are the outermost and innermost radius of the ejecta, respectively.
In our model, the merger ejecta with $M_{\rm ej}=0.01M_{\odot}$, $v_{\rm min}=0.01c$, $v_{\rm max}=0.4c$, and $\delta=1.5$ (e.g., \citealp{Yu2018}) is adopted.

\begin{figure}
\centering
\includegraphics[width=0.6\textwidth]{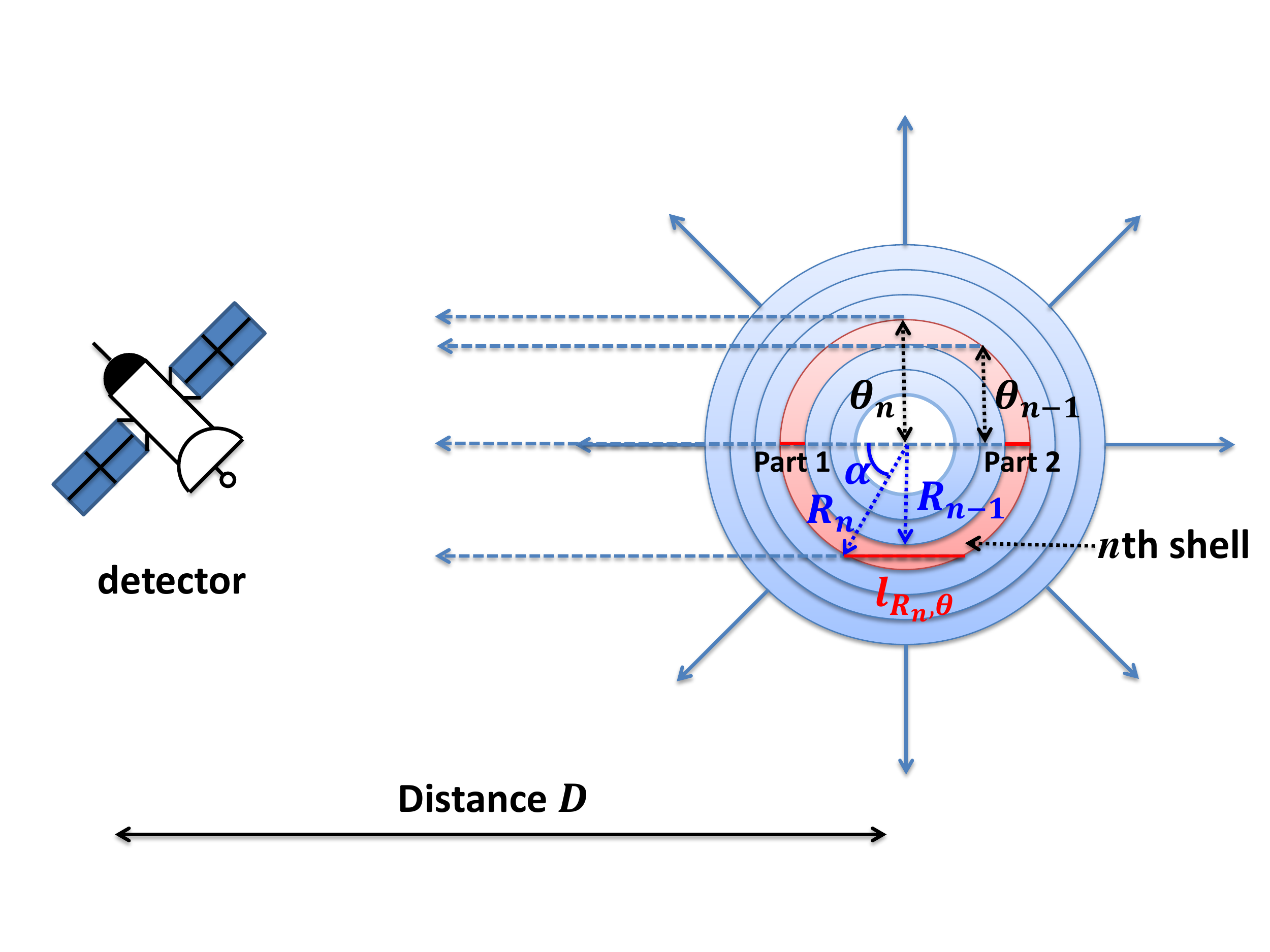}
\caption{Sketch of the $r$-process ejecta model for the neutron star mergers.
The $r$-process material is distributed in a series of spherical shells with different expansion velocity $v_n$.}
\label{sketch}
\end{figure}

A sketch of the ejecta model is illustrated in Figure~\ref{sketch}.
For different viewing angle, the photons will travel in different shells toward the observer.
Similar to \cite{Wang2020}, the angle of the line of sight to the $n$th shell with respect to the line between ejecta centre and detector is $\theta_n\approx {R_n}/D$, where $D$ is the distance between the detector and the source. Since the pathway of emission from different parts of the $n$th shell to the observer is different, our calculation of their emission in different cases are described as below.

\begin{itemize}
\item In the case of $\theta_{n-1}<\theta\leq\theta_n$, the length of the emitting region in the $n$th shell along the line of sight at angle $\theta$ is $l(R_{n}, \theta)=2\sqrt{(R_n)^2-(D\theta)^2}$, and its emission intensity is given by

\begin{equation}
I_{\epsilon}(R_n,\theta) = \int_{0}^{l(R_n,\theta)} j_{\epsilon} e^{-\tau^{\rm in}_{\epsilon}(R_n,\theta)} dl,
\end{equation}
where $\tau^{\rm in}_{\epsilon}(R_n,\theta)$ is the optical depth for photons within the $n$th shell itself,
\begin{equation}
\tau^{\rm in}_{\epsilon}(R_n,\theta)=
\int_{0}^{l(R_n,\theta)} \rho_{\rm ej}(R_n)\kappa(\epsilon) dl.
\end{equation}
Considering the absorption of the outer shells, the observed intensity then is
\begin{equation}
I_{\epsilon}^{\rm obs}(R_n,\theta)=
I_{\epsilon}(R_n,\theta)e^{-\tau^{\rm out}_{\epsilon}(R_n,\theta)},
\end{equation}
where $\tau^{\rm out}_\epsilon (R_n,\theta)$ is the optical depth of the outer shells,
\begin{equation}
\tau^{\rm out}_{\epsilon}(R_n,\theta)=\int_{l(R_{n},\theta)}^{l(R_N,\theta)}\rho_{\rm ej}(R_n)\kappa(\epsilon) dl.
\end{equation}

\item
For case of $0<\theta\leq\theta_{n-1}$, the shell along the light of sight is divided into two separated segments: Part 1 and Part 2. The length of each segment is $l(R_{n}, \theta)=\sqrt{(R_n)^2-(D\theta)^2}-\sqrt{(R_{n-1})^2-(D\theta)^2}$. The emission of Part 1 passes though the outer shell only, being the same as that described in the case of $\theta_{n-1}<\theta\leq\theta_n$. The emission of Part 2 passes through the iner shells and Part 1. The observed emission intensity from Part 1 and Part 2 is given by
\begin{equation}
\label{eq:out}
I_{\epsilon}^{\rm obs}(R_n,\theta)=
I_{\epsilon}(R_n,\theta)e^{-\tau^{\rm out}_{\epsilon}(R_n,\theta)}
+ I_{\epsilon}(R_n,\theta)e^{-\tau^{\prime}_{\epsilon}(R_n,\theta)},
\end{equation}
where $\tau^{'}$ is the total optical depth for photons from Part 2 to the observer,
\begin{equation}
\tau^{\prime}_{\epsilon}(R_n,\theta)=
\begin{cases}
\int_{l(R_{n},\theta)}^{l(R_N,\theta)} \rho_{\rm ej}(R_n)\kappa(\epsilon) dl
+ 2\int_{l(R_{1},\theta)}^{l(R_n,\theta)} \rho_{\rm ej}(R_n)\kappa(\epsilon) dl,~~~(0<\theta\leq\theta_1);\\
\int_{l(R_{n},\theta)}^{l(R_N,\theta)} \rho_{\rm ej}(R_n)\kappa(\epsilon) dl
+ 2\int_{l(R_{m+1},\theta)}^{l(R_n,\theta)} \rho_{\rm ej}(R_n)\kappa(\epsilon) dl
+ \int_{l(R_{m},\theta)}^{l(R_{m+1},\theta)} \rho_{\rm ej}(R_n)\kappa(\epsilon) dl,\\~~~(\theta_1<\theta\leq\theta_{n-1}),
\end{cases}
\end{equation}
where $m$ is the innermost shell that the photon pathway at angle $\theta$ passes through, which is given by  $m=\left[\frac{D\theta-D\theta_1}{(R_N-R_1)/N}\right]$.
\end{itemize}

The observed flux contributed by the $n$th shell can be obtained by
\begin{equation}
F_{\epsilon}^{\rm obs}(R_n)=\int I_{\epsilon}^{\rm obs}(R_n,\theta)\cos\theta d\Omega
= 2\pi \int_{0}^{\theta_n} I_{\epsilon}^{\rm obs}(R_n,\theta)\theta d\theta.
\end{equation}
Thus, the total observed photon flux of the merger ejecta can be obtained by summarizing the contributions of all shells:
\begin{equation}
F_{\epsilon}^{\rm obs}=\sum_n F_{\epsilon}^{\rm obs}(R_n).
\end{equation}

Considering the effect of Doppler shift, the photon energy in the observer frame $\epsilon_{\rm obs}$ is related to that in the rest frame by
\begin{equation}
\epsilon_{\rm obs}=\frac{\epsilon}{\Gamma(1-\beta\cos{\alpha})},
\end{equation}
where $\Gamma$ is the Lorentz factor, $\Gamma=(1-\beta^2)^{-1/2}$, $\beta=v_n/c$, and
$\alpha$ is the angle between the radius vector and the line of sight.

In order to calculate the opacity of the merger ejecta, we take into account four processes of gamma-ray photons in the matter: photoelectric absorption, Compton scattering, pair production, and Rayleigh scattering.
The total opacity of the ejecta is associated with the species of nuclides inside the merger ejecta and their abundance.
In our model, the total opacity is estimated based on the nuclide composition, i.e.,
\begin{equation}
\kappa(\epsilon)=\sum_i A_i Y_i\kappa_i(\epsilon),
\end{equation}
where $\kappa_i(\epsilon)$ is the opacity of the $i$th nuclide and $A_i$ is the corresponding atomic mass number.
The opacity values of the element from hydrogen $(Z=1)$ to fermium $(Z=100)$ are adopted from the XCOM database published by the National Institute of Standards and Technology (NIST) website\footnote{https://www.nist.gov/pml/xcom-photon-cross-sections-database}.

\section{Radioactive Gamma-ray Emission}
\label{sec3}

\subsection{The Abundance Pattern and Gamma-Ray Energy}

\begin{figure}
\centering
\includegraphics[width=0.8\textwidth]{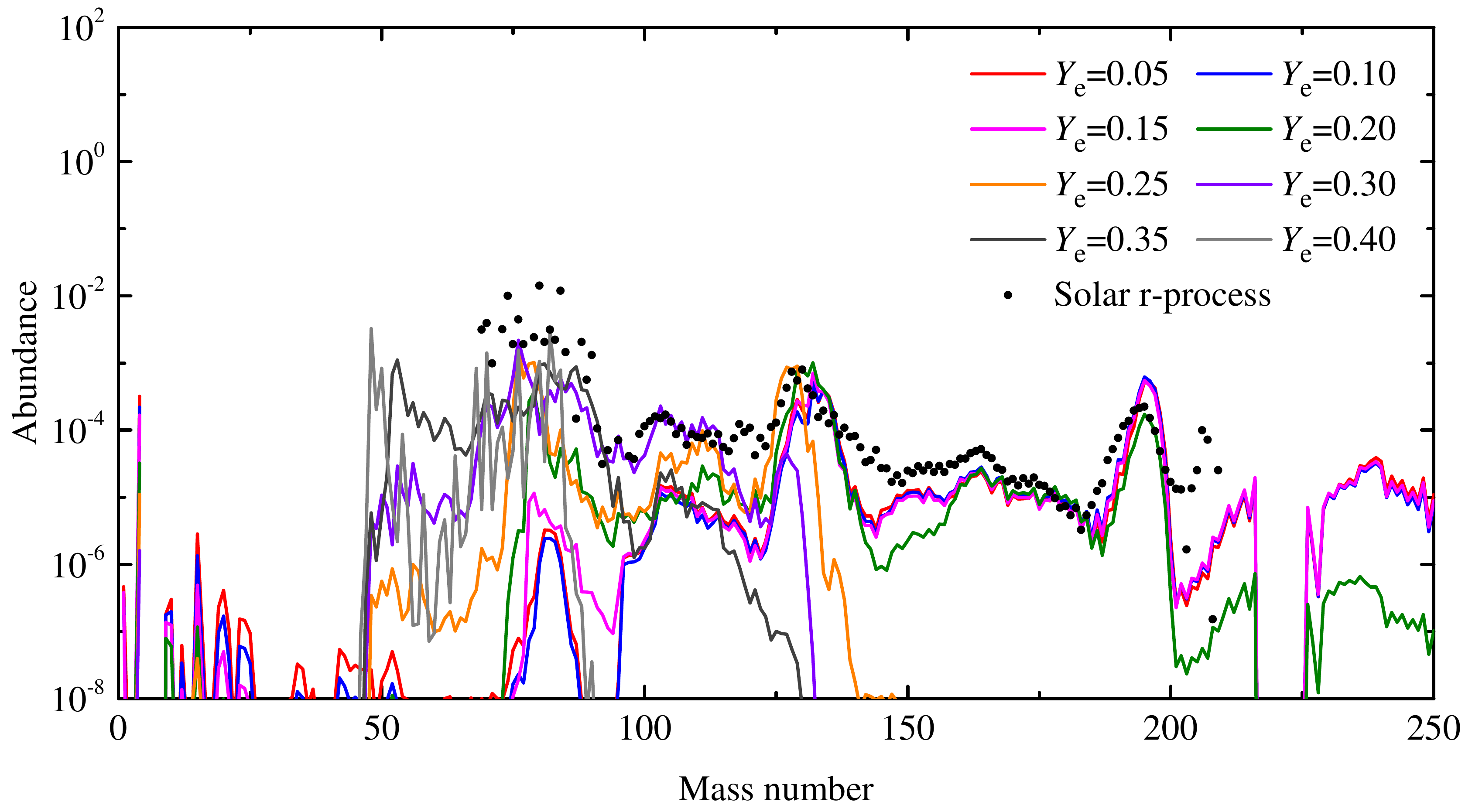}
\caption{The final abundance patterns in the situation with different initial electron fractions $Y_{\rm e}$. The observed solar $r$-process abundances are taken from \cite{Arnould2007}.}
\label{Abundance}
\end{figure}

In Figure~\ref{Abundance}, we show the final abundance patterns
in the situation with different initial electron fractions $Y_{\rm e}$.
For comparison, we also plot the observed solar $r$-process abundances,
which is taken from \cite{Arnould2007}.
For the dynamical ejecta with $Y_{\rm e}\lesssim0.20$, it is in good agreement of the second, third, and rare-Earth peak positions with the solar $r$-process abundances.
The abundance patterns are very similar for the situations with low $Y_{\rm e}$ because these cases are neutron-rich enough to produce nuclides with $A\gtrsim250$.
As the ejecta becomes less neutron-rich, the $r$-process is not fully proceeded because there are not enough neutrons to reach the third $r$-process peak.
This can be found in the wind ejecta with $Y_{\rm e}\gtrsim0.25$,
where the $r$-process fails to reach the third peak,
being instead by producing nuclides around the first $r$-process peak and some iron peak elements.

\begin{figure}
\centering
\includegraphics[width=0.6\textwidth]{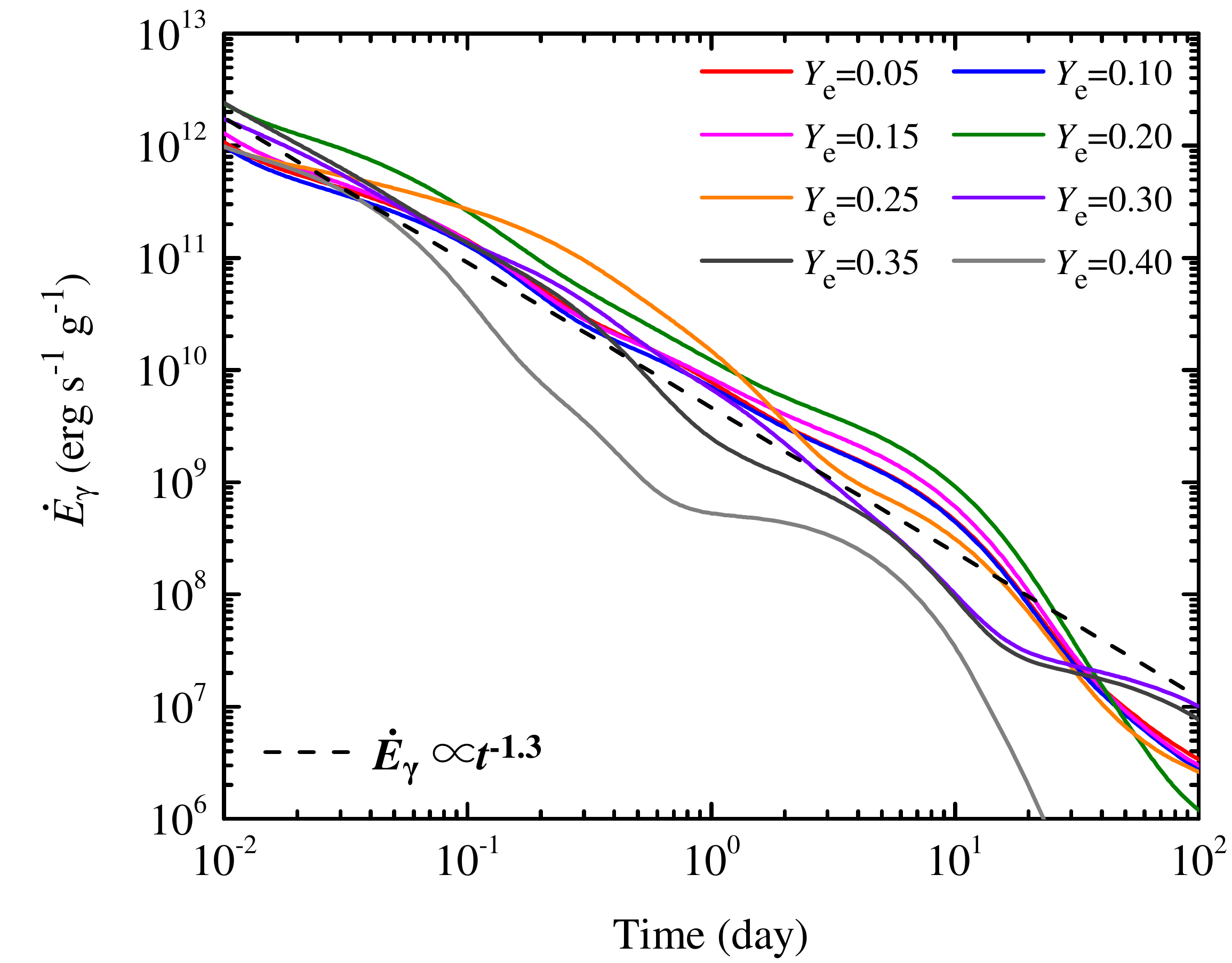}
\caption{The total gamma-ray energy generation rate with different initial electron fractions $Y_{\rm e}$. The dotted line indicates the power-law gamma-ray energy generation rate, i.e., $\dot{E_{\gamma}}\propto t^{-1.3}$.}
\label{heating}
\end{figure}

The gamma-ray energy generation rate with different initial electron fractions $Y_{\rm e}$ are shown in Figure~\ref{heating}, where the dotted line indicates the power-law gamma-ray energy generation rate, i.e., $\dot{E_{\gamma}}\propto t^{-1.3}$.
One can observe that the gamma-ray energy generation rate of merger ejecta can be roughly described with a power-law function.
However, it is changed in the situations with $Y_{\rm e}\gtrsim0.35$.
This is due to that the final composition of the ejecta with $Y_{\rm e}\gtrsim0.35$ is dominated by one or a few individual nuclides, which govern the radioactive  gamma-ray production.

To identify the nuclides which make dominant contribution to the gamma-ray energy,
we calculate the generated gamma-ray energy of each nuclide during its radioactive decay.
In Figure~\ref{dominant1}, we show the dominant nuclides to the gamma-ray energy generation in dynamical ejecta.
The dominant contributions of gamma-ray energy generation come from the nuclides around the second $r$-process peak.
In particular, it is clear that the most important nuclide for the generation of gamma-rays between $1$ day and $10$ days is $^{132}$I.
The gamma-ray energy generation rate of $^{132}$I is higher than the other nuclides by a factor of at least $3-10$ in the dynamical ejecta.
This is owing to that $^{132}$I is largely produced from the decay of doubly magic $^{132}$Sn ($50$ protons and $82$ neutrons) in the $r$-process.
In the decay chain of $^{132}$Sn ($39.7$~s) $\rightarrow$$^{132}$Sb ($2.8$~min) $\rightarrow$$^{132}$Te ($3.2$~d) $\rightarrow$$^{132}$I ($2.3$~h) $\rightarrow$$^{132}$Xe, the corresponding energies of the released gamma-rays are $1.28$~MeV, $2.49$~MeV, $0.21$~MeV, and $2.26$~MeV, respectively.
Then we get a large gamma-ray energy from radioactive $^{132}$I between $1$ day and $10$ days after merger.

We also show the dominant nuclides to the gamma-ray energy generation of wind ejecta in Figure~\ref{dominant2}.
At the wind ejecta with $Y_{\rm e}\gtrsim0.25$, the dominant nuclides contributing to the gamma-ray energy generation rate between $1$ day and $10$ days mainly come from $^{128}$Sb, $^{127}$Sb, $^{77}$Ge, and $^{72}$Ga.
Table~\ref{Edominant} lists the dominant nuclides that contribute to the gamma-ray energy generation rate for different initial electron fractions.

\subsection{The Opacity of $r$-process Elements}

\begin{figure}
\centering
\includegraphics[width=0.6\textwidth]{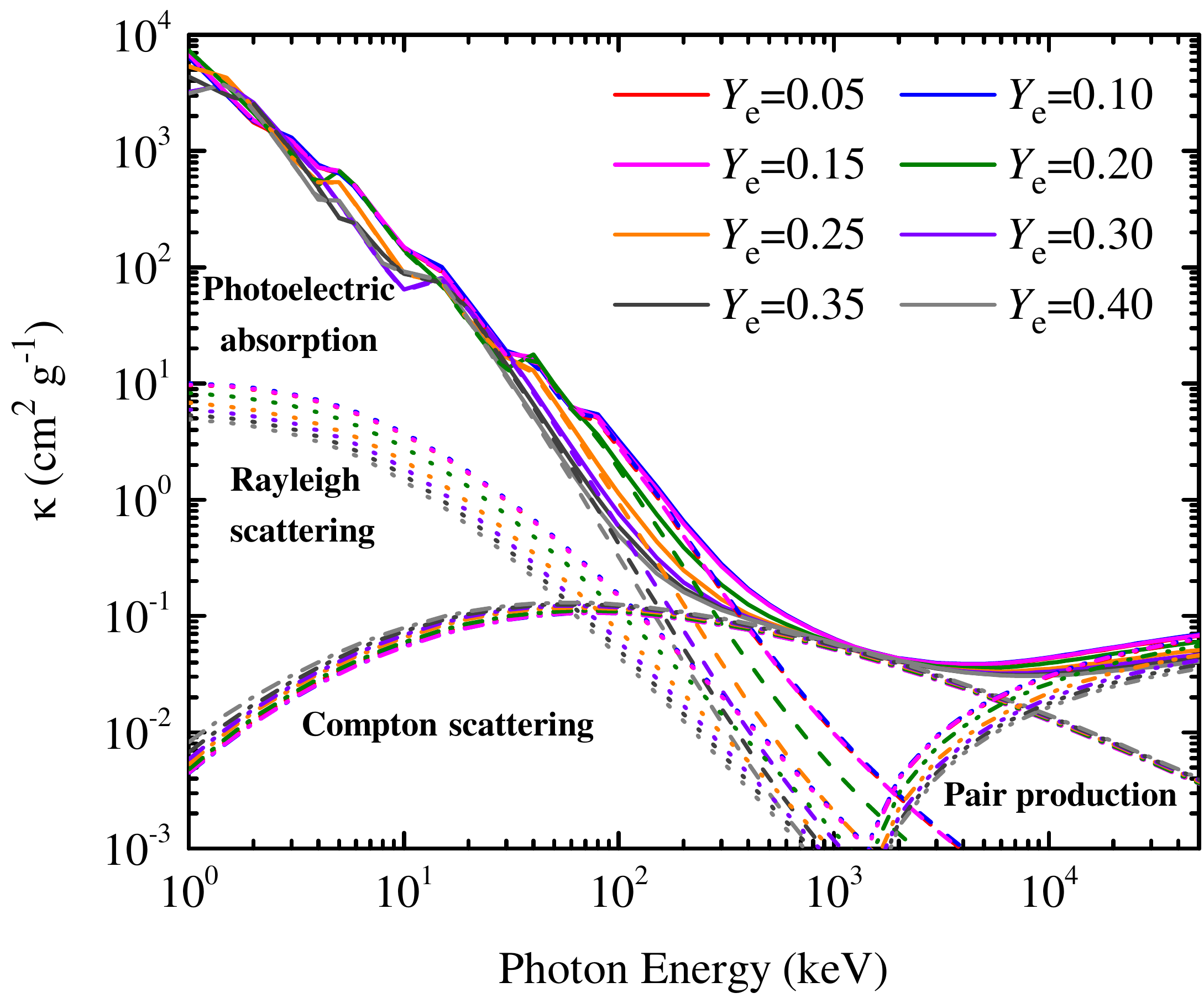}
\caption{The opacities of $r$-process material with different nuclide compositions at $t=1$ day (solid lines). Here, we take into account four processes of gamma-ray photon in the matter: photoelectric absorption (dashed lines), Compton scattering (dash-dotted lines), pair production (double-dash-dotted lines), and Rayleigh scattering (dotted lines).}
\label{kappa}
\end{figure}

To calculate the spectrum of the gamma-ray emission from the merger ejecta, we need to consider the effect of absorption and scattering by ejecta material.
In Figure~\ref{kappa}, we show the total gamma-ray opacity caused by the four processes, including photoelectric absorption, Compton scattering, pair production, and Rayleigh scattering.
It is found that the opacity increases quickly with decreasing photon energy.
For $\epsilon\leq200$~keV, the interaction of gamma-ray photons with matter is dominated by the photoelectric absorption, which is larger than that of the Rayleigh scattering and Compton scattering by orders of magnitude.
For $200$~keV $\leq\epsilon\leq5$~MeV, the opacity is dominated by Compton scattering, while the pair production in the nuclear field become important at energies $\gtrsim5$~MeV.

As shown in Figure~\ref{kappa}, the total opacity in a few hundred keV range is sensitive to the nuclide compositions.
The opacity of heavy material is larger by a factor of $\gtrsim1.5$ than that of lighter elements at low energies $\lesssim500$~keV because photoelectric absorption is enhanced by high-$Z$ atoms.
Therefore, the opacity of the dynamical ejecta can be larger than that of the wind ejecta.

\subsection{The Spectrum of Gamma-Ray Emission}

In Figure~\ref{Flux}, we show the observed gamma-ray spectra with different initial electron fractions $Y_{\rm e}$.
It is found that the observed photon flux in the high-energy range decreases with time as neutron-rich isotopes gradually decay to stability.
The low-energy part of the observed photon flux increases as time goes on, since the photons of energy smaller than a few hundred keV suffer the photoelectric absorption by the atoms in the ejecta during the initial optically thick stage.
The observed photon flux in dynamical ejecta have very similar shapes,
since the nuclide compositions in dynamical ejecta are similar.

As can be found in Figure~\ref{Flux}, the observed gamma-ray spectra have several distinct peaks both in the dynamical ejecta and the wind ejecta.
To identify the radioactive nuclides that make dominant contribution to the spectral peak of gamma-ray spectrum,
we have calculated the contribution of each nuclide to the observed photon flux.
The nuclides we find to be dominant source of gamma-ray spectrum in dynamical ejecta are consistent.
For the spectral peak around $700$~keV, there are several bright gamma-ray lines come from $^{132}$I with energies of $522.65$~keV, $630.19$~keV, $667.71$~keV, $772.60$~keV, and $954.55$~keV, which is about $35\%$ of the total observed photon flux.
The dominant contribution to the spectral peak around $250$~keV comes from $^{132}$Te with energy of $228.16$~keV, which is about $15\%$ of the total observed photon flux.
The low-energy spectral peak around $90$~keV comes from $^{133}$Xe with energy of $81.00$~keV.
The spectral peak near $50$~keV is generated by the radioactive $^{132}$Te.
Note that, when the ejecta is optically thick, the Doppler broadening of emission lines is asymmetric, as only the photons distributing in the near side of the sphere can be seen.

For the observed photon flux in wind ejecta, we see that the gamma-ray spectrum is very sensitive to the initial electron fraction $Y_{\rm e}$.
The dominant contributions to the spectral peak mainly come from $^{128}$Sb, $^{127}$Sb, $^{77}$Ge, $^{73}$Ga, $^{72}$Ga, $^{72}$Zn, and $^{67}$Cu.
At the wind ejecta with $Y_{\rm e}\sim0.25$, there are several bright gamma-ray lines come from $^{128}$Sb and $^{127}$Sb.
For higher initial electron fractions ($Y_{\rm e}\gtrsim0.30$), the spectral peaks of gamma-ray spectrum come from the nuclides around the first $r$-process peak.
The radioactive $^{72}$Ga is responsible for the spectral peaks around $2400$~keV and $850$~keV, which produce several bright gamma-ray lines with eneriges of $2507.72$~keV, $2201.59$~keV, $894.33$~keV, and $834.13$~keV.
The spectral peaks around $250$~keV and $400$~keV mainly come from $^{77}$Ge with energies of $211.03$~keV, $215.51$~keV, $264.45$~keV, and $416.35$~keV.
The spectral peak near $150$~keV is generated by the radioactive $^{72}$Zn.
The low-energy spectral peak around $100$~keV come from $^{67}$Cu with energy of $93.31$~keV.
The dominant nuclides responsible for the spectral peak are listed in Table~\ref{Ndominant}.

In summary, the $r$-process network calculations of the merger ejecta suggest that $^{132}$Te, $^{132}$I, $^{131}$I, $^{133}$Xe, $^{133}$I, $^{128}$Sb, $^{127}$Sb, $^{77}$Ge, $^{73}$Ga, $^{72}$Ga, $^{72}$Zn and $^{67}$Cu are the dominant nuclides contributing to the gamma-ray spectra on a timescale of a few days.
For the dynamical ejecta, the decay chain of $^{132}$Te ($t_{1/2}=3.21$~days) $\rightarrow$ $^{132}$I ($t_{1/2}=0.10$~days) $\rightarrow$ $^{132}$Xe produces several bright gamma-ray lines with energies of $228.16$~keV, $667.71$~keV, and $772.60$~keV.
In the case of the lanthanide free wind ejecta, the decay chain of $^{72}$Zn ($t_{1/2}=1.93$~days) $\rightarrow$ $^{72}$Ga ($t_{1/2}=0.59$~days) $\rightarrow$ $^{72}$Ge also produces several bright gamma-ray lines with energies of $144.70$~keV, $834.13$~keV, $2201.59$~keV, and $2507.72$~keV.
These decay chains would be the promising one to be detected by future observations.

\section{Summary and Discussion}
\label{sec4}

\begin{figure}
\centering
\includegraphics[width=0.8\textwidth]{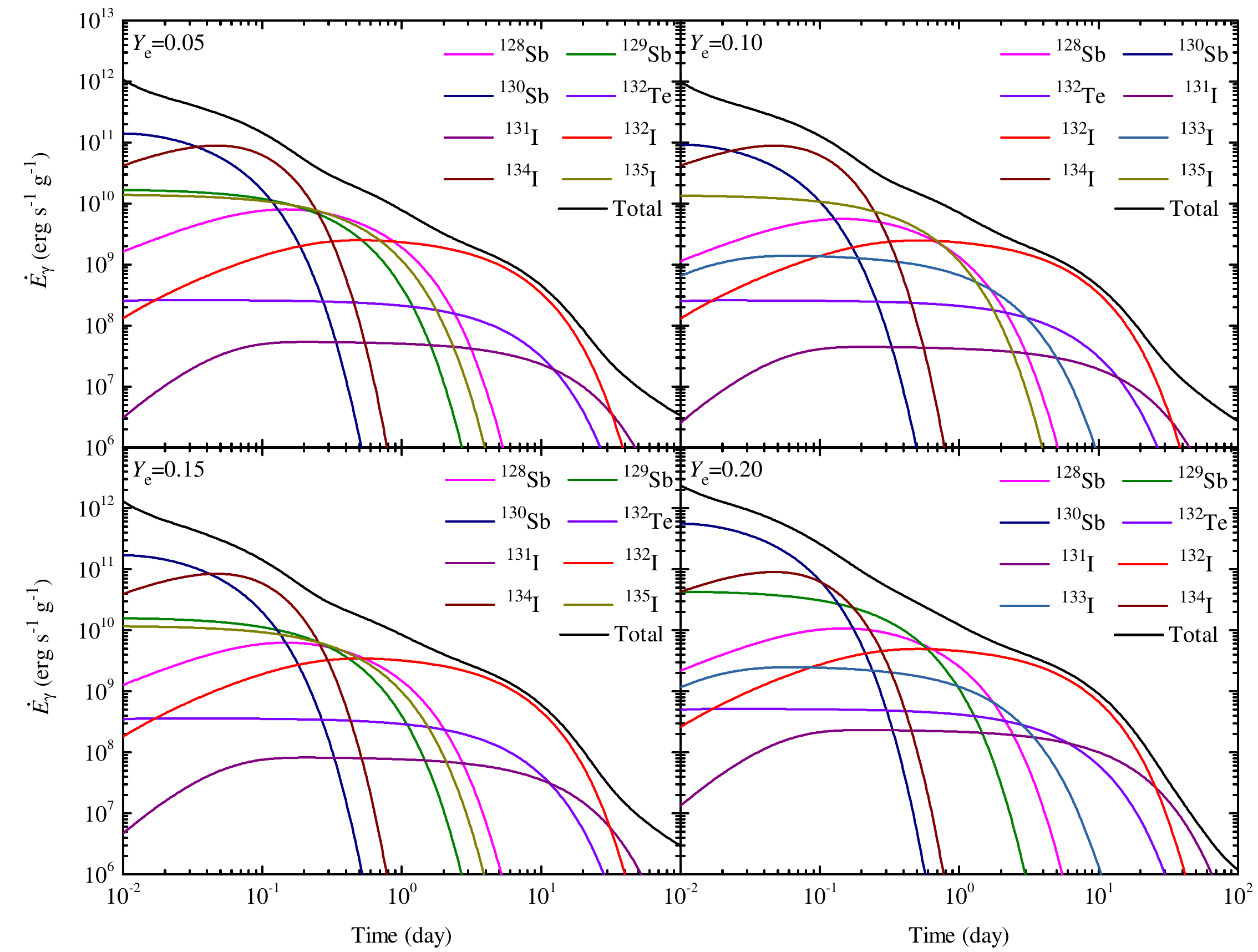}
\caption{The gamma-ray energy generation rate produced by dominant nuclides in dynamical ejecta $(Y_{\rm e}\lesssim0.20)$.}
\label{dominant1}
\end{figure}

\begin{figure}
\centering
\includegraphics[width=0.8\textwidth]{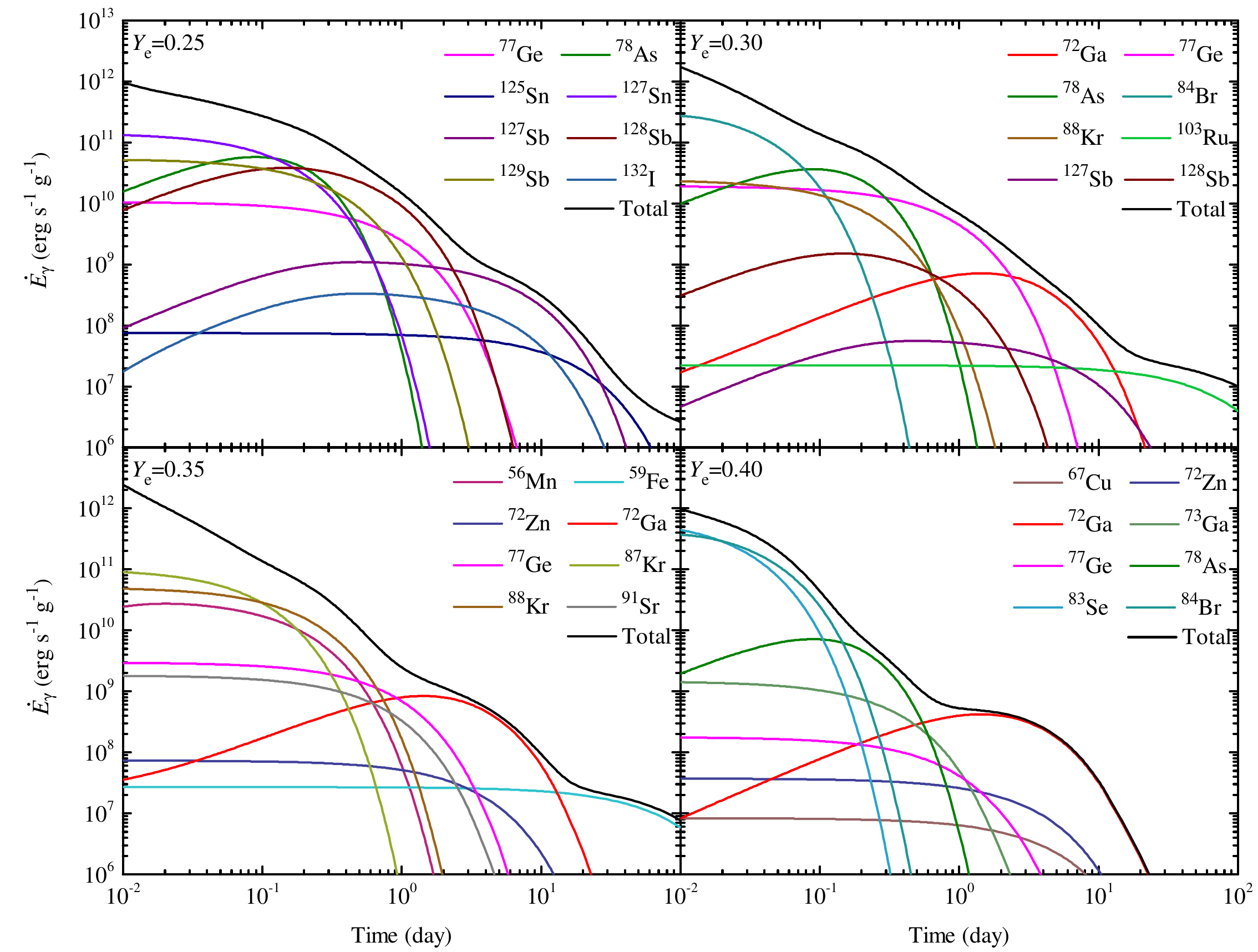}
\caption{The gamma-ray energy generation rate produced by dominant nuclides in wind ejecta $(Y_{\rm e}\gtrsim0.25)$.}
\label{dominant2}
\end{figure}

\begin{figure}
\centering
\includegraphics[width=0.4\textwidth]{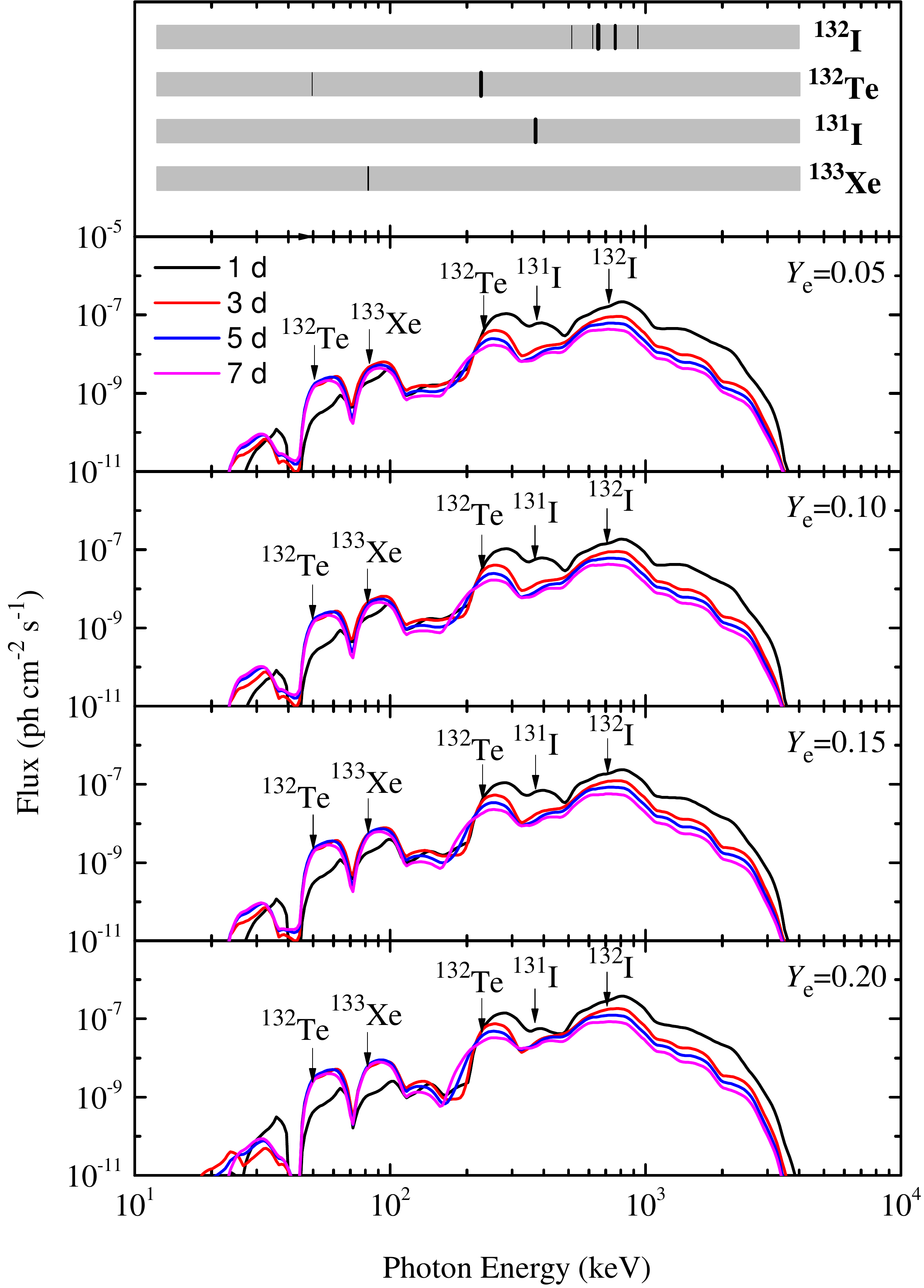}
\includegraphics[width=0.4\textwidth]{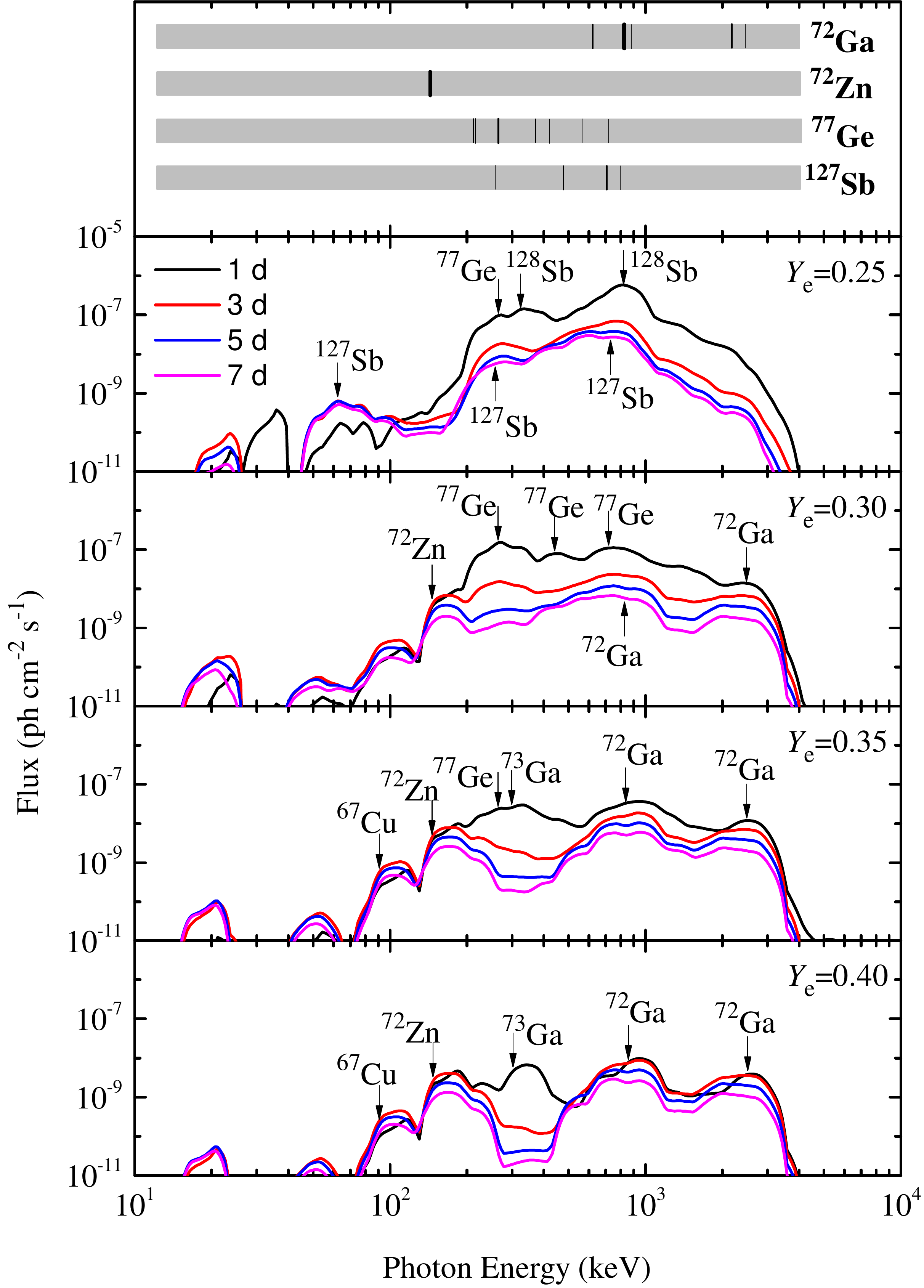}
\caption{The observed spectra of the gamma-ray emission in both the dynamical ejecta $(Y_{\rm e}\lesssim0.20)$ and the wind ejecta $(Y_{\rm e}\gtrsim0.25)$.
Colored lines represent the observed spectra at different time.
The dominant nuclides to the spectral peaks are shown in the figure.}
\label{Flux}
\end{figure}

In this paper, we studied the energy and spectrum of gamma-ray emission produced by the radioactive decay of $r$-process elements freshly synthesized in a neutron star merger ejecta.
Basing on the detailed $r$-process nucleosynthesis calculations
and a multi-shell model for radiative transport,
we calculated the radioactive gamma-ray emission and identified the features in the emission spectrum associated with $r$-process elements.
For the dynamical ejecta in the situation with low initial electron fractions ($Y_{\rm e}\lesssim0.20$),
the dominant contributors of gamma-ray energy are the nuclides around the second $r$-process peak ($A\sim130$) (Figure.~\ref{dominant1}).
The decay chain of $^{132}$Te ($t_{1/2}=3.21$~days) $\rightarrow$ $^{132}$I ($t_{1/2}=0.10$~days) $\rightarrow$ $^{132}$Xe produces several bright gamma-ray lines with energies of $228.16$~keV, $667.71$~keV, and $772.60$~keV (left panel of Figure.~\ref{Flux}), which would be the most promising decay chain to be detected by the MeV gamma-ray detectors.
Our result is consistent with the previous work by \cite{Korobkin2020} on a similar timescale, which also appear spectral peaks around $250$~keV and $700$~keV from the spectra of dynamical ejecta.
The decay chain of $^{132}$Te is also the dominant source of heating rate obtained by \cite{Lippuner2015} and \cite{Zhu2021} with two different $r$-process nucleosynthesis network code.
In the case of wind ejecta with high initial electron fractions ($Y_{\rm e}\gtrsim0.30$), the dominant contributors of gamma-ray energy are the nuclides around the first $r$-process peak ($A\sim80$) (Figure.~\ref{dominant2}).
The decay chain of $^{72}$Zn ($t_{1/2}=1.93$~days) $\rightarrow$ $^{72}$Ga ($t_{1/2}=0.59$~days) $\rightarrow$ $^{72}$Ge produces several bright gamma-ray lines with energies of $144.70$~keV, $834.13$~keV, $2201.59$~keV, and $2507.72$~keV(right panel of Figure.~\ref{Flux}).
This result is similar to those estimated by \cite{Wu2019a}, which suggests that the decay chain of $^{72}$Zn plays a crucial role in powering the kilonova lightcurve of AT~2017gfo.

Our calculations do not include the gamma-ray emission from the fission process being due to lack of radiation data for fissions in the NuDat2 database. This is not significantly affected our results since the fission process only affects the gamma-ray emission above $3.5$~MeV at timescales longer than $10$ days (\citealp{Wang2020}).
In addition, the secondary photons from Compton scattering are ignored in our model, which may affect the gamma-ray spectrum in the low-energy band.
The observed spectra in the low-energy band may become relatively smooth due to the contribution of secondary photons, similar to the results obtained by \cite{Korobkin2020}, but the spectra peaks from the dominant decay chains should still be identified.
Only nuclides whose gamma-ray radiation data are available in the NuDat2 database are included in our calculations of gamma-ray emission from the $r$-process nucleosynthesis.
This may lead to underestimate the gamma-ray emission flux.
However, this cannot significantly affect on the derived gamma-ray spectrum since the gamma-ray spectrum are dominated by the nuclides with long half-life ($t_{1/2}\gtrsim10^4$~s, see Table~\ref{Ndominant}).
These nuclides are close to the valley of stability and their experimental data  are available \citep{Hotokezaka2016,Li2019}.
Note that a simple, symmetric split model for fission reactions is adopted in our $r$-process simulations.
This may overestimate the abundance of nuclides around the second $r$-process peak \citep{Mumpower2018} and the corresponding gamma-ray emissions.

The resulting gamma-ray line fluxes on a timescale of $1-7$ days are $10^{-9}\sim 10^{-7}$~ph~cm$^{-2}$ s$^{-1}$ in the photon energy range of $0.05-3$~MeV at a distance of $40$~Mpc. The sensitivity of the current MeV gamma-ray missions, such as \emph{INTEGRAL} \citep{Diehl2013},
is $10^{-4}\sim 10^{-5}$~ph~cm$^{-2}$ s$^{-1}$ in the MeV band, being much lower than the line fluxes derived from our analysis. The sensitivities of the proposed next-generation missions, such as \emph{AMEGO} (All-sky Medium Energy Gamma-ray Observatory, \citealp{Moiseev2017,Rando2017}), the \emph{e-ASTROGAM} space mission \citep{Tatischeff2016}, \emph{ETCC} (Eletron Tracking Compton Camera, \citealp{Tanimori2015,Tanimori2017}) and \emph{LOX} (Lunar Occultation Explorer, \citealp{Miller2018}), are $\sim10^{-7}-10^{-5}$~ph~cm$^{-2}$~s$^{-1}$. The gamma-ray lines would be marginally detectable with these missions.

\acknowledgments
We thank Bing Zhang, Hou-Jun L$\ddot{\rm u}$, Shan-Qin Wang, Ning Wang, Jia Ren, and Rui-Chong Hu for fruitful discussion and to the anonymous referee for helpful comments.
This work was supported by the National Natural Science Foundation of China (Grant Nos.11533003, 11851304, 11773007, and U1731239) and the Guangxi Science Foundation (Grant Nos.AD17129006, 2017AD22006, 2018GXNSFFA281010, and 2018GXNSFGA281007). Li-Xin Li was supported by the National Natural Science Foundation of China (Grant Nos.11973014 and 11711303).

\newpage

\clearpage
\newpage
\begin{table}
\centering
\caption{Dominant nuclides contributing to the total gamma-ray energy generation rate on the observational timescale of $1$ day and $10$ days.}
\label{Edominant}
\begin{tabular}{ccc}
\hline\hline
Time (day)~~&~~$Y_{\rm e}$~~&~~Isotope~~\\
\hline
~~~1~~~&~~0.05~~&~~~$^{132}$I, $^{128}$Sb, $^{135}$I, $^{133}$I, $^{135}$Xe~~~\\
~~~~~~~&~~0.10~~&~~~$^{132}$I, $^{128}$Sb, $^{135}$I, $^{133}$I, $^{135}$Xe~~~\\
~~~~~~~&~~0.15~~&~~~$^{132}$I, $^{128}$Sb, $^{135}$I, $^{133}$I, $^{135}$Xe~~~\\
~~~~~~~&~~0.20~~&~~~$^{132}$I, $^{128}$Sb, $^{133}$I, $^{135}$I, $^{129}$Sb~~~\\
~~~~~~~&~~0.25~~&~~~$^{128}$Sb, $^{77}$Ge, $^{129}$Sb, $^{127}$Sb, $^{132}$I~~~\\
~~~~~~~&~~0.30~~&~~~$^{77}$Ge, $^{72}$Ga, $^{128}$Sb, $^{112}$Ag, $^{105}$Ru~~~\\
~~~~~~~&~~0.35~~&~~~$^{72}$Ga, $^{77}$Ge, $^{91}$Sr, $^{88}$Kr, $^{73}$Ga~~~\\
~~~~~~~&~~0.40~~&~~~$^{72}$Ga, $^{73}$Ga, $^{77}$Ge, $^{72}$Zn, $^{67}$Cu~~~\\
\hline
~~~10~~&~~0.05~~&~~~$^{132}$I, $^{132}$Te, $^{131}$I, $^{127}$Sb, $^{140}$La~~~\\
~~~~~~~&~~0.10~~&~~~$^{132}$I, $^{132}$Te, $^{131}$I, $^{127}$Sb, $^{140}$La~~~\\
~~~~~~~&~~0.15~~&~~~$^{132}$I, $^{132}$Te, $^{131}$I, $^{127}$Sb, $^{211}$Bi~~~\\
~~~~~~~&~~0.20~~&~~~$^{132}$I, $^{131}$I, $^{132}$Te, $^{127}$Sb, $^{140}$La~~~\\
~~~~~~~&~~0.25~~&~~~$^{127}$Sb, $^{132}$I, $^{125}$Sn, $^{131}$I, $^{132}$Te~~~\\
~~~~~~~&~~0.30~~&~~~$^{72}$Ga, $^{103}$Ru, $^{127}$Sb, $^{95}$Zr, $^{125}$Sn~~~\\
~~~~~~~&~~0.35~~&~~~$^{72}$Ga, $^{59}$Fe, $^{72}$Zn, $^{103}$Ru, $^{95}$Zr~~~\\
~~~~~~~&~~0.40~~&~~~$^{72}$Ga, $^{72}$Zn, $^{67}$Cu, $^{66}$Cu, $^{59}$Fe~~~\\
\hline
\end{tabular}\\[1mm]
\end{table}

\begin{table}
\centering
\caption{Dominant nuclides contributing to the spectral peaks between $1$~day and $1$~week.}
\label{Ndominant}
\begin{tabular}{ccccc}
\hline\hline
$Y_{\rm e}$~~&~~Isotope~~&~~~$t_{1/2}$~(s)~~~&~~Line Energy (keV)~~&~~Intensity~~\\
\hline
0.05-0.20 & $^{132}$I & $8.26\times10^3$ & 667.71 & 98.70$\%$ \\
~ & $^{132}$I & $8.26\times10^3$ & 772.60 & 75.60$\%$ \\
~ & $^{132}$I & $8.26\times10^3$ & 954.55 & 17.60$\%$ \\
~ & $^{132}$I & $8.26\times10^3$ & 522.65 & 16.00$\%$ \\
~ & $^{132}$I & $8.26\times10^3$ & 630.19 & 13.30$\%$ \\
~ & $^{132}$Te & $2.77\times10^5$ & 228.16 & 88.00$\%$ \\
~ & $^{132}$Te & $2.77\times10^5$ & 49.72 & 15.00$\%$ \\
~ & $^{133}$Xe & $4.53\times10^5$ & 81.00 & 36.9$\%$ \\
~ & $^{131}$I & $6.93\times10^5$ & 364.49 & 81.50$\%$ \\
~ & $^{133}$I & $7.50\times10^4$ & 529.87 & 87.00$\%$ \\
\hline
0.25-0.40 & $^{72}$Ga & $5.08\times10^4$ & 834.13 & 95.45$\%$ \\
~ & $^{72}$Ga & $5.08\times10^4$ & 2201.59 & 26.87$\%$ \\
~ & $^{72}$Ga & $5.08\times10^4$ & 629.97 & 26.13$\%$ \\
~ & $^{72}$Ga & $5.08\times10^4$ & 2507.72 & 13.33$\%$ \\
~ & $^{72}$Ga & $5.08\times10^4$ & 894.33 & 10.14$\%$ \\
~ & $^{72}$Zn & $1.67\times10^5$ & 144.70 & 82.78$\%$ \\
~ & $^{77}$Ge & $4.04\times10^4$ & 264.45 & 53.30$\%$ \\
~ & $^{77}$Ge & $4.04\times10^4$ & 211.03 & 30.00$\%$ \\
~ & $^{77}$Ge & $4.04\times10^4$ & 215.51 & 27.90$\%$ \\
~ & $^{77}$Ge & $4.04\times10^4$ & 416.35 & 22.70$\%$ \\
~ & $^{77}$Ge & $4.04\times10^4$ & 714.37 & 7.50$\%$ \\
~ & $^{73}$Ga & $1.75\times10^4$ & 297.32 & 78.80$\%$ \\
~ & $^{67}$Cu & $2.23\times10^5$ & 93.31 & 16.10$\%$ \\
~ & $^{127}$Sb & $3.33\times10^5$ & 685.70 & 36.80$\%$ \\
~ & $^{127}$Sb & $3.33\times10^5$ & 473.00 & 25.80$\%$ \\
~ & $^{127}$Sb & $3.33\times10^5$ & 783.70 & 15.10$\%$ \\
~ & $^{127}$Sb & $3.33\times10^5$ & 252.40 & 8.50$\%$ \\
~ & $^{128}$Sb & $3.26\times10^4$ & 754.00 & 100.00$\%$ \\
~ & $^{128}$Sb & $3.26\times10^4$ & 743.30 & 100.00$\%$ \\
~ & $^{128}$Sb & $3.26\times10^4$ & 314.10 & 61.00$\%$ \\
\hline
\end{tabular}\\[1mm]
\end{table}

\end{document}